# Single-Layer CrI$_3$ Grown by Molecular Beam Epitaxy

Peigen Li,[§,#] Cong Wang,[∥,#] Jihai Zhang,[§] Shenwei Chen,[§] Donghui Guo,[§] Wei Ji,[∥,*] and Dingyong Zhong[§,‡,*]

[§]*School of Physics, Sun Yat-sen University, 510275 Guangzhou, China*

[∥]*Beijing Key Laboratory of Optoelectronic Functional Materials & Micro-Nano Devices, Department of Physics, Renmin University of China, 100872 Beijing, China*

[‡]*State Key Laboratory for Optoelectronic Materials and Technologies, Sun Yat-sen University, 510275 Guangzhou, China*

[#]*The authors equally contributed to this work.*

[*]*Corresponding authors: wji@ruc.edu.cn (theoretical part); dyzhong@mail.sysu.edu.cn*

**Abstract**

Single- and few-layer chromium triiodide (CrI$_3$), which has been intensively investigated as a promising platform for two-dimensional magnetism, was usually prepared by mechanical exfoliation. Here, we report on the growth of single-layer CrI$_3$ by molecular beam epitaxy under ultrahigh vacuum. The atomic structures and local density of states have been revealed by scanning tunneling microscopy (STM). Iodine trimers, each of which consists of three I atoms surrounding a three-fold Cr honeycomb center, have been identified as the basic units of the topmost I layer. Different superstructures of single-layer CrI$_3$ with characteristic periodicity around 2-4 nm were obtained on Au(111), but only pristine structure was observed on graphite. At elevated temperatures (423 K), CrI$_3$ was partially decomposed, resulting in the formation of single-layer chromium diiodide. Our bias-dependent STM images suggest that the unoccupied and occupied states are distributed spatial-separately, which is consistent with our density functional theory calculations. The effect of charge distribution on the superexchange interaction in single-layer CrI$_3$ was discussed.

**Keywords**: two-dimensional magnetic materials, CrI$_3$, molecular beam epitaxy, scanning tunneling microscopy

Two-dimensional (2D) magnetic materials, which serve as ideal platforms for investigating spin-related emergent phenomena at reduced dimensions and exhibit potential applications in optoelectronics and spintronics,[1,2] have long been sought. However, as revealed by Mermin-Wagner theorem, long-range magnetic order in 2D systems with isotropic local magnetic interactions are thermodynamically instable at finite temperatures owing to enhanced thermal fluctuations.[3] By introducing magnetic anisotropy, on the other hand, 2D magnetism may exist in a number of layered van der Waals (vdW) materials, for example, transition metal chacogenides and halides.[4-6] Among them, single- and bi-layer chromium triiodide ($CrI_3$) has been recently confirmed with intralayer ferromagnetism (FM) and interlayer antiferromagnetism (AFM).[7,8] The FM coupling in single-layer $CrI_3$, which exhibits a Curie temperature of 45 K, slightly lower than that of the bulk counterpart (61 K), is mainly ascribed to the superexchange interaction between the neighboring Cr atoms mediated by the bridging I atoms.[9] The easy magnetization direction is out of plane with an anisotropy energy about 0.50-0.69 meV/Cr, according to density functional theory (DFT) calculations.[10,11] The magnetic tunneling junction consisting of four-layer $CrI_3$ sandwiched with two graphene layers as the electrodes performs a magnetoresistance change as large as 19,000%,[12] demonstrating the advantages of layered vdW materials for fabricating advanced spintronic devices.[12-15] Besides $CrI_3$, other layered materials, such as $Cr_2Ge_2Te_6$, $CrCl_3$, $CrBr_3$, $Fe_3GeTe_2$ and $FePS_3$, have been already obtained and verified with long-range magnetic order.[16-21]

So far, mechanical exfoliation is the major method for preparing 2D magnetic materials. High-quality 2D materials with reduced structural defects can be prepared by mechanical exfoliation, but the size is usually limited in the micrometer range. Chemical vapor deposition has also been used for the growth of single-crystalline nanoplates of various layered magnetic materials.[22] As a matured film growth technique, molecular beam epitaxy (MBE) has been widely used in fundamental research and industry for preparing high-quality films and heterostructures with atomic-level structural controllability. By combining scanning tunneling microscopy (STM), local structural and magnetic properties of MBE-prepared 2D magnetic materials can be in-situ investigated at the atomic level.[23,24] Here, we investigate single-layer

CrI$_3$ on Au(111) and graphite substrates grown by MBE under ultrahigh vacuum. The surface structures of epitaxial CrI$_3$ have been revealed with atomic resolution for the first time by STM. According to our STM images and DFT calculations, we found that the distribution of the unoccupied states near the Fermi energy at the single-layer CrI$_3$ surface is concentrated upon the three-fold Cr honeycomb centers, while the occupied states are mainly located upon the Cr-Cr bridge sites. In addition, we have also successfully prepared single-layer chromium diiodide (CrI$_2$) after partially removing iodine from CrI$_3$ at elevated temperatures.

**Results and discussion**

Figure 1a shows the large-scale STM image of single-layer CrI$_3$ on the Au(111) surface with sub-monolayer coverage. The size of the CrI$_3$ islands, which are initially nucleated at the step edges of the Au(111) surface, is in the range of several tens nm up to 100 nm with an irregular or proximately round shape. At sub-monolayer coverage, the areas without CrI$_3$ are covered with a layer of iodine. Samples with a complete CrI$_3$ monolayer were also prepared (Figure S1). X-ray photoelectron spectroscopy (XPS) measurement on the sample with monolayer coverage indicates a Cr: I ratio close to 1: 4.4 (Table S1), implying the existence of an iodine buffer layer between CrI$_3$ and Au(111). The apparent height of the single-layer CrI$_3$ islands is about 6.8 Å (see inset of Figure 1a), consistent with the thickness of the samples prepared by mechanical exfoliation (7 Å).[7] In the zoom-in image (Figure 1b), a superstructure of hexagonal lattice appears with a height oscillation about 30 pm, superposing on the periodic atomic structure of CrI$_3$. Figure 1c is the Fourier transform image of Figure 1b, containing three sets of hexagonal patterns. The spots marked by yellow, red and white circles correspond to the approximately closed-packed iodine atoms of the topmost layer, the Bravais lattice of CrI$_3$ and the superstructure, respectively. The superstructure exhibits a periodicity of 2.93 nm, 4.26 times larger than the CrI$_3$ lattice constant (6.867 Å),[25] and is rotated by 18.6° with respect to the CrI$_3$ lattice. The relationship between the supercell and the primitive CrI$_3$ lattice can be described as $\begin{pmatrix} \boldsymbol{a}_{\text{dot}} \\ \boldsymbol{b}_{\text{dot}} \end{pmatrix} = \begin{pmatrix} 4.83 & -1.57 \\ 1.57 & 3.26 \end{pmatrix} \begin{pmatrix} \boldsymbol{a}_{CrI_3} \\ \boldsymbol{b}_{CrI_3} \end{pmatrix}$. It was observed that in some islands more than one domain of such "dot" superstructure exist with domain boundaries in between. As shown in Figure 1d, there are two domains separated by the domain boundaries marked by white dashed lines. The superstructures in the two domains have their orientations (red and blue arrows) about 19° rotated with each other. However, the whole CrI$_3$ lattice (green arrows) is continuous at the domain boundaries without line defects (Figure S2). The atomic-resolution STM images of single-layer CrI$_3$ are shown in Figure 1e and 1f. The bright protrusions exhibit a nearly close-packed hexagonal lattice, corresponding to individual I atoms from the topmost layer. However, they are not uniformly arranged, resulting in an unambiguous feature of the triangle-shaped I trimer, in which three I atoms are slightly aggregated together. The I trimers form a larger hexagonal lattice with a periodicity of 6.95 Å, consistent with the lattice constant of the CrI$_3$ layer from previous experiments and calculations.[11,25] Given the symmetry and lattice constant, one can deduce that the three I atoms in each trimer are surrounding a honeycomb center of the underlying Cr layer, as shown by the structural model superposed on the STM images in Figure 1f.

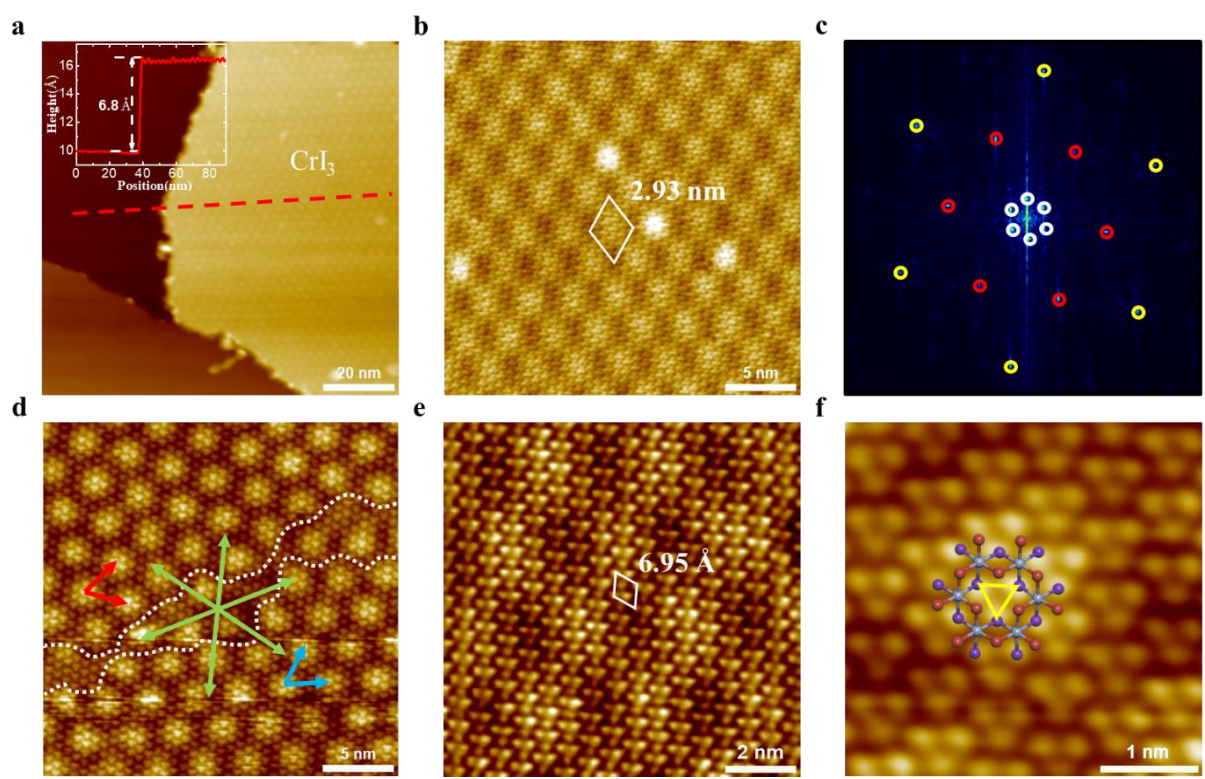

**Figure 1.** Dot phase of single-layer CrI$_3$ on Au(111). (a) Large-scale STM image ($V_s$ = 2 V, I = 100 pA). The inset is a height profile along the red dashed line. (b) The superstructure with periodicity of 2.93 nm ($V_s$ = 0.9 V, I = 100 pA). (c) Fourier transform map of (b). The hexagonal patterns marked by yellow, red and white circles correspond to the periodicities of topmost I atoms, CrI$_3$ and superstructure, respectively. (d) The domain boundary of the dot phase ($V_s$ = 2 V,

I = 100 pA). The domain boundaries are marked by white dashed lines. Red and blue arrows represent the orientations of the superstructures, while green arrows show the orientation of the continuous CrI$_3$ lattice. (e) Atomic-resolution STM image of CrI$_3$ (V$_s$ = 0.7 V, I = 100 pA). The white rhombus represents the unit cell with $a$ = 6.95 Å. (f) Atomic-resolution STM image of CrI$_3$ superposed with the structure model (V$_s$ = 0.7 V, I = 100 pA). Purple, red and gray balls represent upper I, bottom I and Cr atoms. Yellow triangle represents the I trimer.

Besides the dominant dot phase described above, minor superstructures of single-layer CrI$_3$ have been also observed on Au(111). As shown in Figure 2a, three different phases coexist: the dot phase, the stripe phase and the pristine phase. Note that all of them appear with the feature of I trimers. Figure 2b and 2c show zoomed STM images of the pristine and stripe phases, respectively. No superstructure exists in the pristine CrI$_3$ layer, while in the stripe phase, a rectangle supercell (1.83 nm by 1.88 nm) appears with the basic vectors written as $\begin{pmatrix} \mathbf{a}_{\text{stripe}} \\ \mathbf{b}_{\text{stripe}} \end{pmatrix} = \begin{pmatrix} 3.08 & 1.54 \\ 0 & 2.74 \end{pmatrix} \begin{pmatrix} \mathbf{a}_{CrI_3} \\ \mathbf{b}_{CrI_3} \end{pmatrix}$, containing about 8 primitive unit cells. By annealing at 433 K for 30 min, a new stripe phase (type II stripe phase) was observed, as shown in Figure 2d and 2e. The type II phase exhibits an even larger parallelogram supercell (3.83 by 2.32 nm, 100°) with the basic vectors written as $\begin{pmatrix} \mathbf{a}_{\text{type II}} \\ \mathbf{b}_{\text{type II}} \end{pmatrix} = \begin{pmatrix} 4.55 & 1.67 \\ -2.24 & 3.89 \end{pmatrix} \begin{pmatrix} \mathbf{a}_{CrI_3} \\ \mathbf{b}_{CrI_3} \end{pmatrix}$, containing about 21 primitive unit cells.

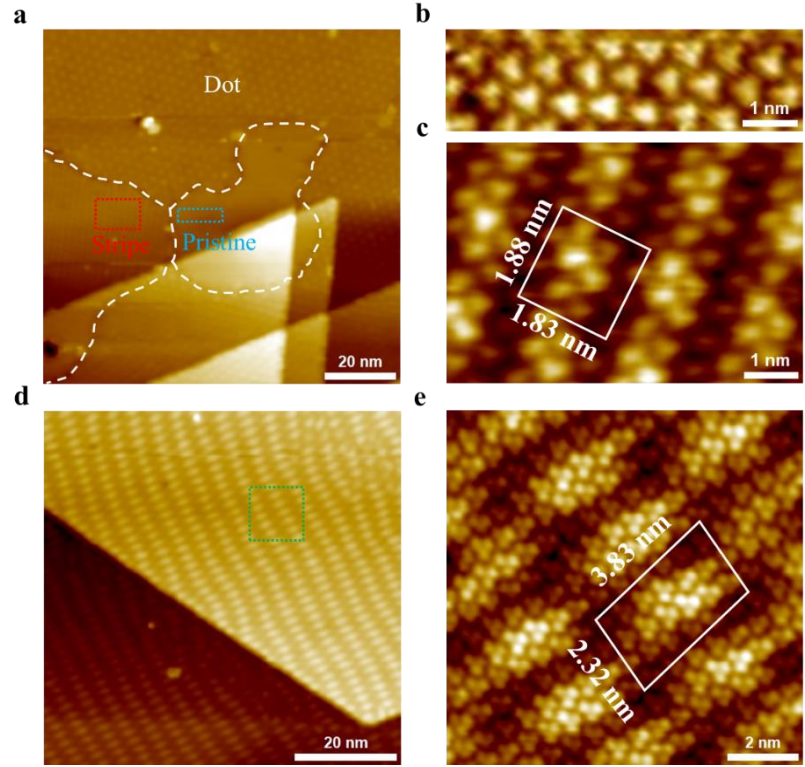

**Figure 2.** Different superstructures of single-layer CrI$_3$ on Au(111). (a) Large-scale STM image of dot phase, stripe phase and pristine phase coexisting in a CrI$_3$ island (V$_s$ = 1.5 V, I = 50 pA). (b) Zoomed STM image of the pristine phase in the area marked by the blue dashed frame in (a) (V$_s$ = 1.4 V, I = 50 pA). (c) Zoomed STM image of the stripe phase in the area marked by the red dashed frame in (a) (V$_s$ = 1.5 V, I = 60 pA). (d) The STM image of the type II stripe phase (V$_s$ = 1.5 V, I = 50 pA). (e) Zoomed STM image of the type II stripe phase in the area marked by the green dashed frame in (d) (V$_s$ = 0.7 V, I = 100 pA).

We have also found that further treating the samples at elevated temperatures (> 423 K) can result in the partial decomposition of CrI$_3$, forming single-layer chromium diiodide (CrI$_2$) consequently. Single-layer CrI$_2$ has been previously predicted to be a likely AFM order and Type-II Dirac cone.[4] Figure 3 shows the STM images measured on the sample after further annealing at 423 K for 30 min. A zigzag phase coexists with the dot phase in Figure 3a. It should be noted that such zigzag phase can also be obtained by growing the sample at higher temperatures (450 K) (see Supplementary Figure S3). The apparent height of the zigzag phase is 0.34 Å smaller than that of the dot phase (inset of Figure 3a). Figure 3b and 3c show the atomic structure of the zigzag phase, indicating that the atoms of the topmost layer are close-packed with a hexagonal lattice (lattice constant 4.08 Å). No periodicity around 6.867 Å and no typical I trimers appear in the zigzag phase, ruling out the formation of a new $CrI_3$ superstructure. On the contrary, the observed lattice constant, 4.08 Å, is well consistent with the layered material $CrI_2$, which have an in-plane lattice constant of 3.92-3.99 Å,[26,27] The reduced thickness compared with $CrI_3$ also imply the formation of single-layer $CrI_2$ (interlayer distance, 6.77 Å).[26,27] Figure 3c shows the atomic-resolution STM image of single-layer $CrI_2$, from which one can find that the topmost atoms in two symmetric directions is arranged uniformly, while the atomic arrangement in the third symmetric direction is wavy. This implies that the zigzag superstructure may originate from the mismatch of the $CrI_2$ lattice with the underlying Au(111) surface. If we further anneal the sample to 456 K, $CrI_3$ will be completely converted into $CrI_2$. Annealing to 482 K, chromium iodide films will be finally decomposed and only ordered iodine adlayer and disordered clusters (probably residual Cr) are left on the surface. The iodine adlayer exhibits a periodicity of 5 Å, which is the same as the iodine-adsorbed Au(111) surfaces reported previously.[28]

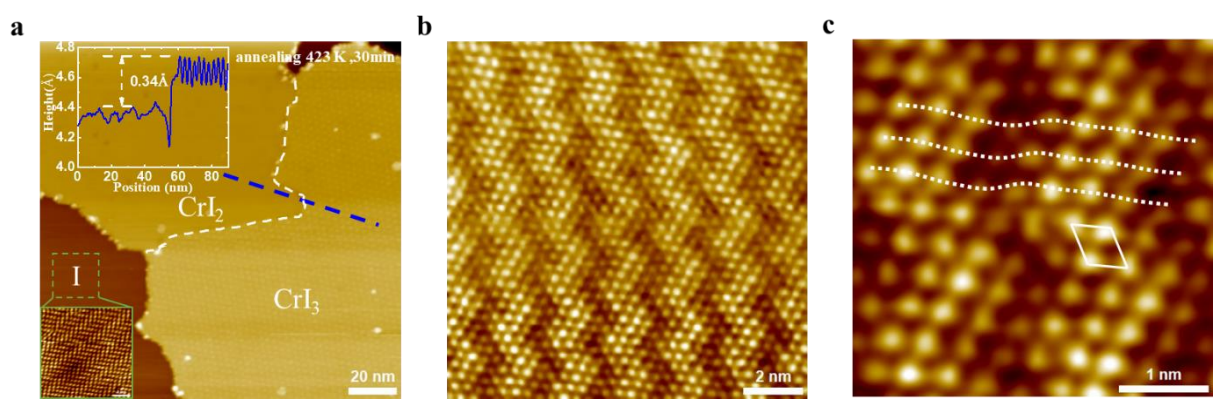

**Figure 3**. Zigzag phase of single-layer $CrI_2$ on Au(111). (a) Coexistence of $CrI_2$ (zigzag phase) and $CrI_3$ (dot phase) after annealing at 423 K for 30 min ($V_s$ = 2.3 V, I = 20 pA). The top inset is a height profile along the blue dashed line. The bottom inset shows the adsorbed I monolayer on Au(111). (b) Zoomed STM image of $CrI_2$ ($V_s$ = 0.5 V, I = 400 pA). (c) Atomic-resolution STM image of $CrI_2$ ($V_s$ = 0.5 V, I = 400 pA). White dashed lines indicate the arrangement of the topmost I layer. A white diamond represents the unit cell with $a$ = 4.08 Å.

Similar growth procedure has been done on graphite substrates to obtain single-layer of both $CrI_2$ and $CrI_3$ (see experimental details). Figure 4a shows the large-scale STM image of single-layer CrI$_2$ islands on graphite. The islands are strip-shaped with an apparent height of 7.5 Å (sample bias, 3.3 V) (Figure S4). The atomic-resolution STM image of single-layer CrI$_2$ on graphite is shown in Figure 4b. Different from the zigzag phase of CrI$_2$ on Au(111), an apparent (1×6) stripe superstructure was observed, superposing the hexagonal atomic lattice with a lattice constant of 3.98 Å, consistent with the intralayer lattice constant of bulk CrI$_2$.[26,27] Due to the inertness of graphite surfaces, it is difficult to obtain an excessive iodine environment, which is important for growing CrI$_3$ and avoiding the formation of CrI$_2$. Therefore, a two-step procedure, *i.e.*, low-temperature codeposition followed by annealing at elevated temperatures, is used to obtain CrI$_3$ on graphite (see experimental details). CrI$_3$ islands with fractal shape are shown in Figure 4c. Increasing I and Cr deposition results in larger domain size (Figure S5). The measured height of the islands is about 7.9 Å (sample bias, 4 V) (Figure S4), which is slightly larger than that of the CrI$_3$ islands (6.8 Å) on Au(111), probably due to the higher density of states (DOS) on Au(111) surfaces. Similar feature of I trimers with lattice constant of 6.84 Å was also observed, as shown in Figure 4d. Detailed structural parameters were listed in Supplementary Table S2. In comparison to the case of Au(111), there is only the pristine structure of CrI$_3$ without any superstructures.

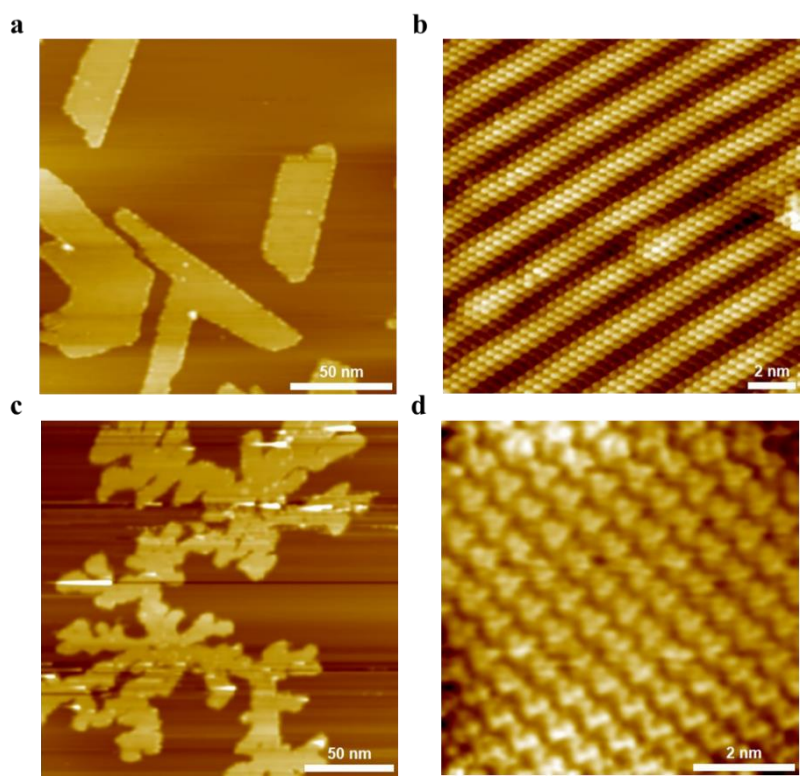

**Figure 4.** Single-layer CrI$_2$ and CrI$_3$ on graphite. (a) Large-scale STM image of CrI$_2$ ($V_s$ = 3.3 V,

I = 20 pA). (b) Atomic-resolution STM image of $CrI_2$ ($V_s$ = 0.7 V, I = 100 pA), showing a stripe superstructure of (1×6) with respect to the pristine $CrI_2$. (c) Large-scale STM image of $CrI_3$ ($V_s$ = 4 V, I = 40 pA). (d) Atomic-resolution STM image $CrI_3$ (Vs = 1 V, I = 100 pA).

Our experimental result indicates the substrate effect on the formation of superstructures in single-layer $CrI_3$. The possible formation mechanisms may relate to, for example, the surface reconstruction of substrates, the intrinsic structural transition in the $CrI_3$ layer itself, or the lattice mismatch between $CrI_3$ and underlying I layer (Moiré pattern). On certain epitaxial layers, such as NaCl grown on Au(111),[29] the Au(111) herringbone reconstruction is still visible. However, the superstructures of $CrI_3$ observed in our experiments is quite dissimilar to the Au(111) herringbone pattern. It has been reported that iodine adlayers on Au(111) surfaces exhibit different structures.[28] The formation of different superstructures in single-layer $CrI_3$ on Au(111) is likely to be caused by different iodine atomic arrangement of the buffer layer. Whether the magnetic properties in single-layer $CrI_3$ are modulated by the superstructures is an open question.

We have carried out bias-dependent STM imaging on single-layer $CrI_3$, unveiling the spatial distribution of occupied and unoccupied DOS near the Fermi energy. Figure 5a and 5c show the STM images of single-layer $CrI_3$ at the bias voltage ($V_s$) of 1 V and −1 V, respectively. At $V_s$ = 1 V, the image displays the typical feature of I trimers, indicating that the unoccupied local density of states (LDOS) of the topmost I layer is distributed upon the Cr honeycomb centers (marked by yellow triangles). When $V_s$ switches to −1 V, the sites upon the Cr honeycomb centers change from bright to dark, while the surrounding edges become brighter (see Figure 5c), appearing with dark triangular holes (marked by black triangles). Such bias-dependent contrast difference was observed repeatedly when switching the bias between 1 and −1 V during tip scanning (Figure 5e and Figure S6), implying that it is not related to the accidental change of the tip electronic states. Importantly, our simulated STM images for both positive and negative biases well resemble the experiment (Figure 5b and 5d). Our results indicate that the distribution of the unoccupied states at the surface of single-layer $CrI_3$ is concentrated upon the Cr honeycomb centers, while the occupied states are mainly located upon the Cr-Cr bridging sites. This assessment to the different distributions of electronic states was also supported by STM images measured with small bias voltages, *i.e.* $V_s$ = –0.1 V and 0.1 V (see Figure S7), where the features of the I trimers and dark triangular holes disappear.

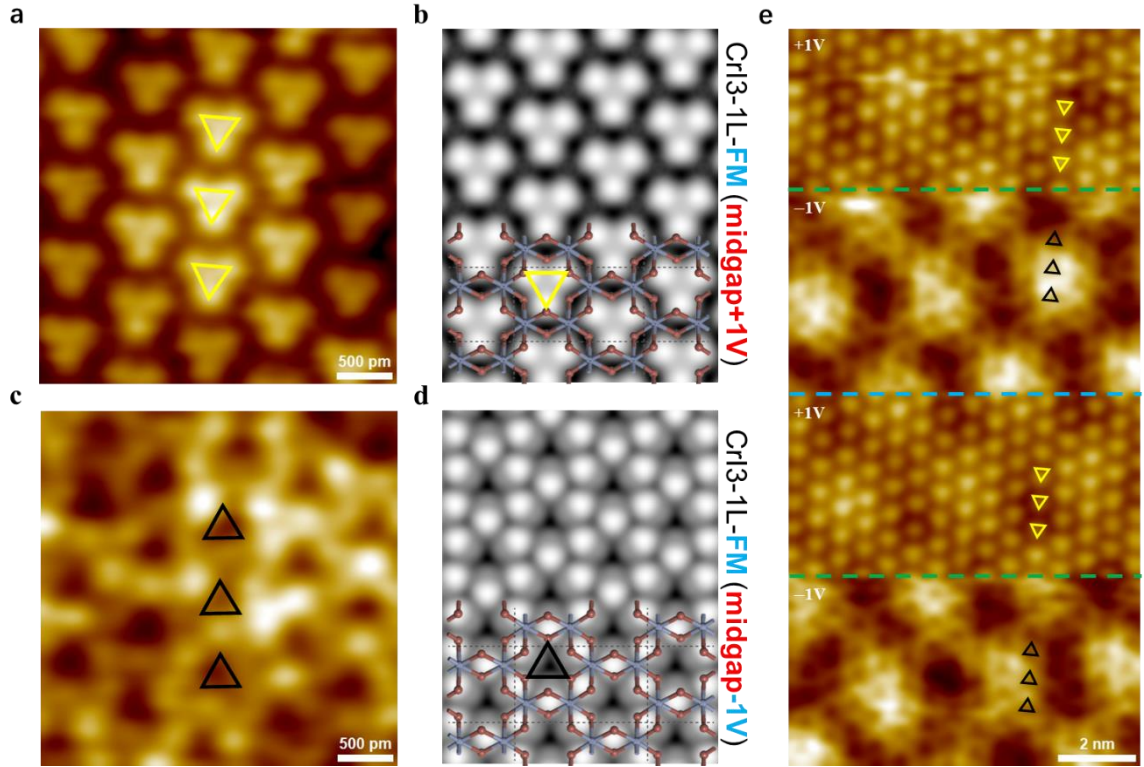

**Figure 5**. Distribution of unoccupied and occupied states at single-layer CrI$_3$ surface. (a) and (b) Experimental and simulated STM images at $V_s$ = 1 V (I = 300 pA), showing the brighter I trimer feature upon the Cr honeycomb centers. (c), (d) Experimental and simulated STM images at $V_s$ = –1 V (I = –200 pA), showing dark triangular holes upon the Cr honeycomb centers. (e) STM image with repeated bias switching between 1 and –1 V. Green dashed lines represent $V_s$ switching from 1 to –1 V while the blue dashed line from –1 V to 1 V. Yellow and black triangles correspond to the I trimers and dark triangular holes upon the Cr honeycomb centers.

DFT calculations have been conducted to investigate the hybridization between Cr and I atoms and the states near the Fermi level, which substantially helps for further understanding the magnetic coupling mechanism of single-layer CrI$_3$. Figure 6a shows a perspective view of the atomic differential charge density of monolayer CrI$_3$ in the FM order, in which pink and green isosurface contours correspond to the electron accumulation and reduction, respectively. Here, we define the Cr-I bond direction as the *x* and *y* axis while the direction perpendicular to the Cr-I plane as the *z* axis (black arrows in Figure 6a). We found that significant charge transfer occurs from Cr $e_g$ orbitals to the bonding regions of Cr-I covalent σ bonds, suggesting that the Cr $e_g$ orbitals bond with I *p* orbitals. Meanwhile, there are three Cr *d* electrons filling in the Cr $t_{2g}$ states with parallel spin orientation, which shows a local magnetic moment of 3.28 $\mu_B$ on each Cr atom. Decomposed spin-polarized band structures shown in Supplementary Figure S8 also confirm this picture. While spin-up Cr $t_{2g}$ states exhibit the local moments, the valence band near the bandgap of CrI$_3$ is mainly comprised of spin-down I $p_z$ orbitals and the conduction band is primarily contributed by the spin-up Cr-$e_g$ and I-$p_x/p_y$ states.

Such band-decomposition is highly consistent with the spin charge density plotted in Figure 6b where the spin up density (red) is mostly around Cr atoms with a minor portion at the $p_z$ states of adjacent I atoms, while the opposite spin density (blue) is observable for I-$p_x/p_y$ orbitals. This particularly spatial distribution of valence and conduction states were indeed observed in STM images. Since STM is a surface sensitive measuring technique, we focus on the I trimer of the topmost layer. The conduction band states, detectable in positive biases, is around I $p_{x/y}$ states (denoted by red arrows). These states extend toward the vacuum and the Cr honeycomb center, giving rise to a bright and defuse trimer-like appearance in the positively biased STM images (Figure 5a and 5b). In terms of negatively biased images, the valence band states mainly reside at the I $p_z$ orbitals (marked by blue arrows). The I $p_z$ orbital points to the center of two neighboring Cr atoms being bridged by this I atom, exhibiting a bright protrusion sitting at the Cr-Cr center, as depicted in Figure 5c and 5d.

The electronic structure of monolayer CrI$_3$ indicates an FM super-exchange mechanism between Cr atoms bridged by two intermediated I atoms with a Cr-I-Cr bond angle of 93° (denoted as J$_1$). Cation Cr$^{3+}$ takes a sp$^3$d$^2$ hybridization where six orbitals orient towards the corners of the Cr-I octahedrons, with 3/2 *e* filled in each orbital and each I-*p* orbital is filled by 5/3 *e*. The super-exchange interaction involves four Cr orbitals from two Cr cations and four I $p_{x/y}$ orbitals from two bridging I atoms, which are totally filled with 38/3 *e* (6 *e* from Cr and 20/3 *e* from I), as illustrated in Figure 6c. Such filling motif offers roughly three channels for virtual hopping of spin-up electrons sitting on Cr sites. Moreover, the I $p_z$ orbital also plays a certain role in enhancing the FM super-exchange as that a portion of spin-up electrons does transfer to the I $p_z$ orbital (Figure 6a and 6b), which may open an additional exchange chancel for the FM interaction.

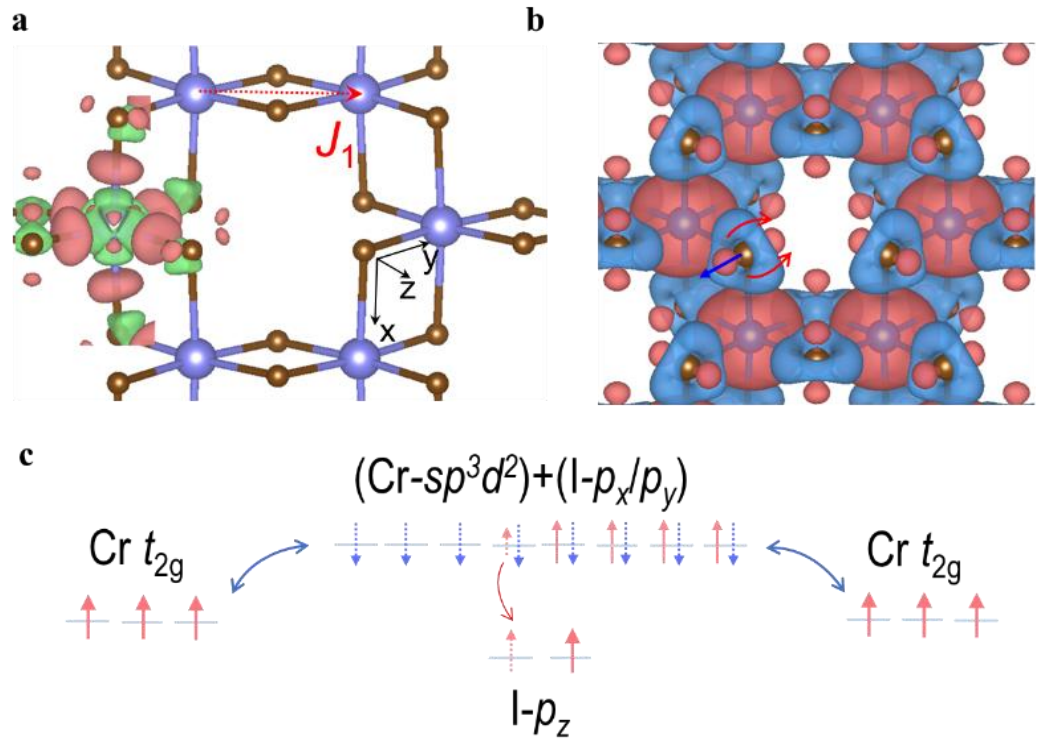

**Figure 6.** Electronic structure of single-layer $CrI_3$ with FM order. (a) Atomic differential charge density. Pink and green isosurfaces correspond to the charge accumulation and reduction after Cr and I atoms bonding together, respectively. (b) Top view of spin charge density of single-layer $CrI_3$ with FM order. Red and blue isosurfaces correspond to the charge with spin component up and down, respectively. (c) Schematic of the FM super-exchange mechanism for nearest spin-exchange bridged by two I atoms (denoted $J_1$), in which red up-oriented and blue down-oriented arrows represent electrons with different spins.

**Conclusions**

In summary, we have successfully synthesized single-layer $CrI_3$ on Au(111) and graphite by MBE under UHV. I trimer is identified as the basic unit of the topmost I layer. Different superstructures of single-layer $CrI_3$ on Au(111) have been obtained. On graphite, single-layer $CrI_3$ with pristine structure has been obtained. We have also prepared single-layer $CrI_2$ on Au(111) by annealing the $CrI_3$ samples or directly growing at elevated temperatures. Through experiments and theoretical simulation, we have unveiled that the different appearance of unoccupied and occupied STM images are ascribed to particularly spatial distributions of $p_{x/y}$ and $p_z$ orbitals of the topmost I layer. A detailed FM super-exchange mechanism of the Cr-I-Cr sites has been unveiled, which helps to understand the strong in-plane FM and highlights the role of I $p_z$ orbitals in tuning magnetism of single- and few-layer $CrI_3$.

**Experimental and calculation details**

**Experimental details.** Single-layer $CrI_3$ and $CrI_2$ were fabricated on Au(111) and highly oriented pyrolytic graphite (HOPG) (Mateck GmbH) substrates under ultrahigh vacuum (UHV) condition with a base pressure of $8\times10^{-10}$ mbar. Au(111) surfaces were cleaned by cycles of sputtering and annealing under UHV condition. HOPG was cleaved in air and immediately transferred into the UHV chamber, followed by annealing to 600 K for degassing. $CrI_3$ powder (purity 90 %), which is decomposed with iodine release at 495 K under UHV condition, was used as the iodine source. Chromium powder (purity 99.996 %, Alfa Aesar) were evaporated from a Knudsen cell at 1223 K. Single-layer $CrI_3$ films with sub-monolayer coverage were prepared by codeposition of chromium and iodine for 30 min onto Au(111) surfaces, which were kept at 425 K. $CrI_2$ films were prepared by codeposition of chromium and iodine for 30 minutes onto Au(111) substrates which were kept at 450 K. In the case of HOPG substrates, due to the relatively low sticking coefficient for iodine at elevated temperatures, a two-step procedure was adopted: codeposition of chromium and iodine at room temperature for 40 min followed by annealing to 463 K. In order to compensate the desorption of iodine from the HOPG surfaces, the iodine source was kept running during the whole annealing process. $CrI_2$ films were grown by codeposition of chromium and iodine for 60 min onto the HOPG substrates which were kept at 463 K. The samples were transferred into the analysis chamber for STM measurement. All STM experiments were performed with constant current mode at 77 K using a chemically etched tungsten tip. $CrI_3$ on Au(111) with monolayer coverage produced on the STM preparation chamber was directly transferred within 2 minutes into the Thermo Fisher XPS with a UHV load-lock and immediately pumped. All spectra were measured in the multi-chamber UHV system equipped with a monochromatic Al Kα X-ray source (1486.6eV).

**Calculation details** Our density functional theory calculations were performed using the generalized gradient approximation and the projector augmented wave method[30] as implemented in the Vienna *ab-initio* simulation package (VASP).[31] A uniform 15×15×1 Monkhorst-Pack *k* mesh was adopted for integration over the first Brillouin zone. A plane-wave kinetic energy cutoff of 700 eV was used during structural relaxations. A sufficiently large distance of $c > 20$ Å along the out-of-plane direction of $CrI_3$ was adopted to eliminate interactions among image layers. Dispersion correction was made at the van der Waals density functional (vdW-DF) level,[32-34] with the optB86b functional for the exchange potential, which was proved to be accurate in describing the structural properties of layered materials.[35-40] During structural relaxation, all atoms and the shape of the supercell were allowed to relax until the residual force per atom was less than 0.01 eV/Å. On-site Coulomb interaction to the Cr *d* orbitals was self-consistently calculated based on a linear-response method,[41] which gives $U = 3.9$ eV and $J = 1.1$ eV, as adopted in a previous calculation.[42]

**Notes**

The authors declared no competing financial interest.

**Acknowledgement**

This work was financially supported by National Natural Science Foundation of China (11574403, 11974431, 11832019, 11622437, 61674171 and 11974422), Guangzhou Science and Technology Project (201707020002), the Strategic Priority Research Program of Chinese Academy of Sciences (XDB30000000), the Fundamental Research Funds for the Central Universities, China, and the Research Funds of Renmin University of China (16XNLQ01). C. W. was supported by the Outstanding Innovative Talents Cultivation Funded Programs 2017 of Renmin University of China. Calculations were performed at the Physics Lab of High-Performance Computing of Renmin University of China and Shanghai Supercomputer Center.

**Supporting information**

XPS of CrI$_3$ on Au(111) with monolayer coverage, The domain boundary of the dot phase, the zigzag phase of single-layer CrI$_2$ on Au(111), the apparent height of single-layer CrI$_2$ and CrI$_3$ on HOPG, detailed structural parameters of the single-layer CrI$_2$ and CrI$_3$, the bias-dependent STM images of single-layer CrI$_3$ surface, the electronic band structure of the FM single-layer CrI$_3$ with spin component of up and down.